\documentclass[a4paper,11pt]{article}
\pdfoutput=1
\usepackage{float}
\usepackage{jheppub}
\usepackage{amsmath,amssymb}
\usepackage{xspace}
\usepackage{color}
\usepackage{hyperref}
\usepackage{epsfig}
\usepackage{xcolor}
\usepackage{xspace}
\usepackage{verbatim}
\usepackage{graphicx}
\usepackage{mathtools}
\usepackage[numbers]{natbib}

\begin{document}
\title{Diffractive and photon-induced production of top quark}

\author[1,2]{Michael Pitt}

\affiliation[1]{Ben-Gurion University of the Negev, Department of Physics,
  Beer-Sheva, Israel }
\affiliation[2]{The University of Kansas, Department of Physics,
Lawrence, USA }

\emailAdd{michael.pitt@cern.ch}

\abstract{The top quark plays a central role in particle physics, as many experiments at the Large Hadron Collider scrutinize its properties within the Standard Model. Although most of the measurements of the top quarks today concentrate on production modes initiated by quarks or gluons, this review will highlight the lesser-explored modes initiated by pomerons or photons. It aims to provide an in-depth look into both the phenomenological studies and the existing experimental measurements, emphasizing the necessity of exploring the diffractive and photon-induced production of top quarks to enhance the accuracy of top-quark measurements.}

\maketitle


\section{Introduction}

Since its discovery in 1995 at Fermilab’s Tevatron collider, the top quark has continued to receive significant attention, as it is a promising pathway for investigating the SM by conducting high-precision measurements of its properties and potentially unveiling new laws of physics by discerning any deviation from SM predictions.

\begin{figure}[H]
\includegraphics[width=0.19\textwidth]{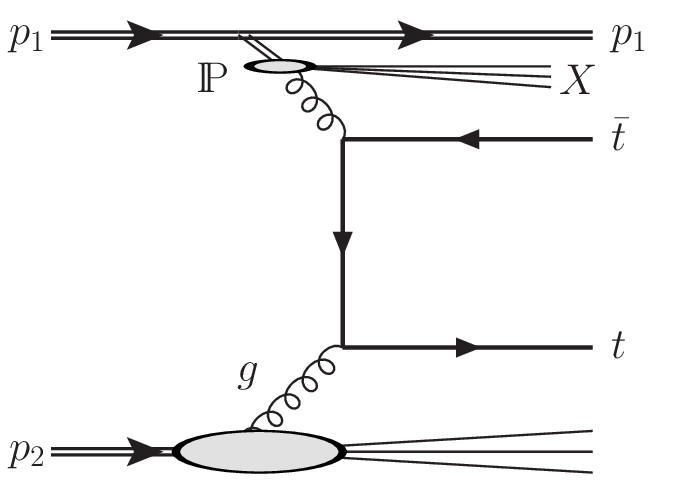}
\includegraphics[width=0.19\textwidth]{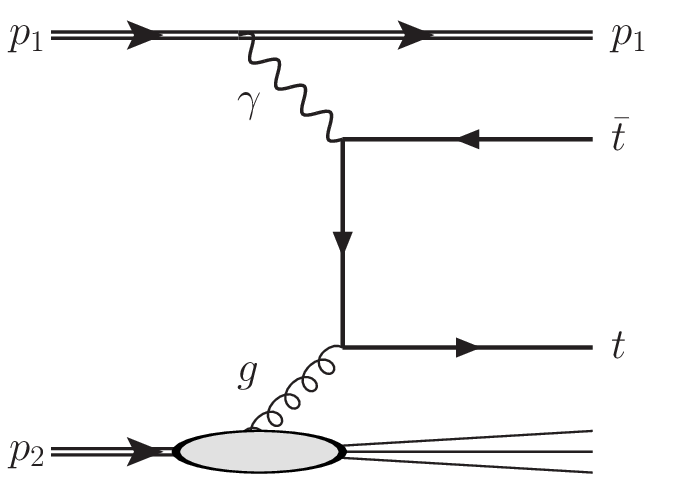}
\includegraphics[width=0.19\textwidth]{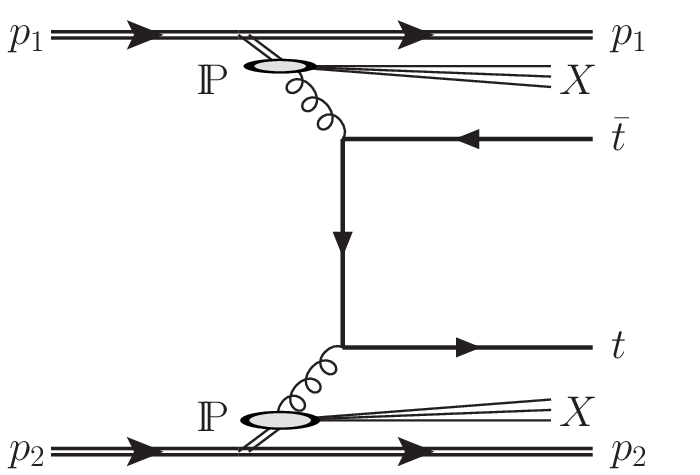}
\includegraphics[width=0.19\textwidth]{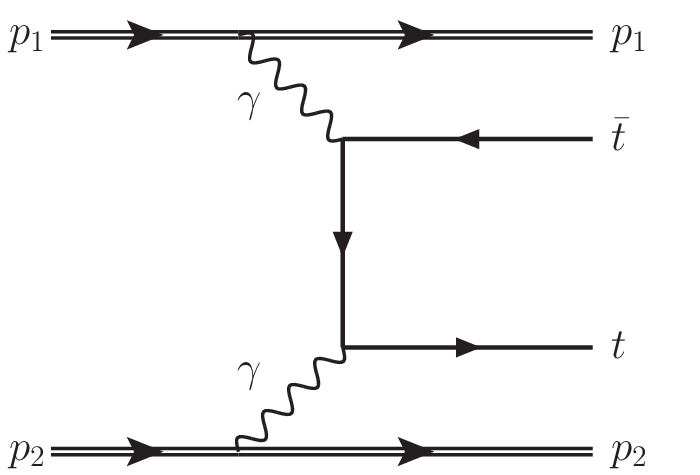}
\includegraphics[width=0.19\textwidth]{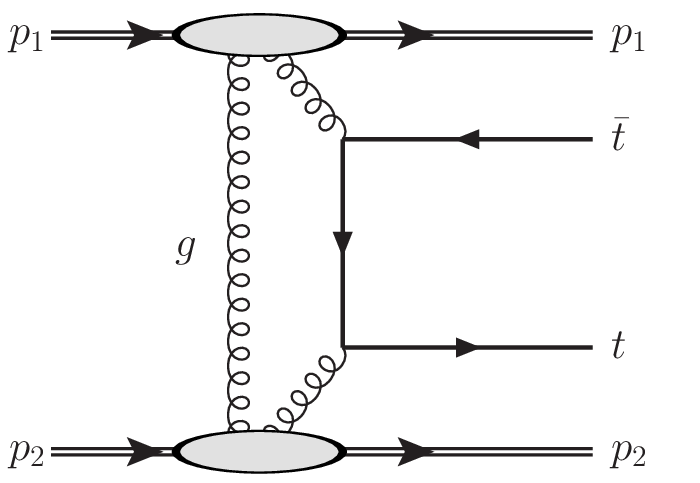}
\caption{Categories of top quark pair production involving color-neutral particles. From left to right: single-diffractive production, photoproduction, double pomeron exchange, central exclusive production via photon exchange, and central exclusive production via pomeron exchange.\label{fig:diagrams}}
\end{figure}

The present measurements of top quarks produced from the collisions of protons at the LHC predominantly focus on production modes initiated by quarks or gluons. However, the production of top quarks initiated by color-neutral particles remains largely unexplored and is the subject of this review. Production modes involving color-neutral particles are typically categorized according to the type of particle involved. For example, top quark pairs can also be produced through photon ($\gamma$) or pomeron ($I\!P$) exchange, and the different production categories are depicted in Figure~\ref{fig:diagrams}.

The photon is an elementary particle responsible for electromagnetic interactions. On the other hand, the pomeron is an object that can be viewed as a color-singlet multi-gluon state, introduced in the early 1960s to describe high-energy hadron-scattering amplitudes (an in-depth review can be found in reference~\cite{Donnachie:2002en}). Over time, with deeper insights into diffractive processes, it became evident that the pomeron has an internal structure that includes quarks as well. Consequently, any process initiated by quarks or gluons can be generated by exchanging pomerons. 

The discussion of the diffractive production of top quarks predates its actual discovery~\cite{Collins_1984} yet, up to the present day, it has lacked experimental verification. The production of top quarks through color-neutral particles can contribute to the total inclusive cross-section on a level of up to a few percent. The events resulting from a photon or pomeron exchange exhibit a distinct radiation pattern, often characterized by an absence of hadronic activity in certain regions of pseudorapidity, known as large rapidity gaps. Additionally, these events may be identified by the presence of an intact proton. The following article will review the theory and experimental data available on these rare production modes.

\section{Monte Carlo event generators}

Diffractive and exclusive processes have been incorporated into various Monte Carlo (MC) event generators. Different MC generators may implement these processes differently based on their underlying physics models. However, they all adhere to the factorization principle~\cite{Collins:1989gx}, where the production cross-section of a system X in hard proton--proton scattering can be viewed as follows: 

\begin{equation}
\sigma\left(\text{pp}\to\text{X}\right) = \sum_\text{i,j} \int \text{dx}_\text{i} f_\text{i}(\text{x}_\text{i},\mu) \int \text{dx}_\text{j}  f_\text{j}(\text{x}_\text{j},\mu) \cdot \sigma_\text{hard}\left(\text{ij}\to\text{X}\right),
\end{equation}

where i, j label the partons that initiate hard scattering, $\sigma_\text{hard}\left(\text{ij}\to\text{X}\right)$ is the parton level cross-section computed perturbatively in terms of powers of $\alpha_S(\mu)$, $\mu$ is the energy scale of the process, and $f(\text{x},\mu)$ are the parton distribution functions (PDFs) of the colliding particles. For color-neutral interaction, these parton densities are often replaced by the following:

\begin{equation}
\text{dx} f(\text{x},\mu) =\int \text{d}t  \int \text{d}\beta \text{I\!F}(\xi,t) f_d(\beta,\mu),
\end{equation}

where $ \text{I\!F}(\xi,t)$ is the flux of color-neutral mediators emitted by a proton as a function of longitudinal momentum ($\xi$) and the momentum transfer ($t$). Here, $\beta = \text{x}/\xi$ and $f_d(\beta,\mu)$ is the parton density function of the color-neutral object (technically related to the diffractive PDF --- dPDF). In the case of direct photon/pomeron exchange processes, whether they are exclusive or semi-exclusive, the value of $f_d(\beta,\mu)$ is set to 1, but the flux will incorporate additional form-factors, denoted by $S^2(\text{b})$, which are derived in terms of impact parameter and their computation varies across different event generators.

The pomeron flux $\text{I\!F}_{\rm{I\!P}/p}\left(\xi,k_T\right)$ is the probability that a pomeron with a given value of $\xi$ and $t$ couples to the proton. Several different parametrizations have been proposed over the years~\cite{Ingelman:1984ns,donnachie1988hard}. The most recent empirical expression was suggested by the H1 \mbox{collaboration~\cite{H1:2006zyl}:}

\begin{equation}
\text{I\!F}_{\rm{I\!P}/p}\left(\xi,t\right) = \xi^{1-2\alpha(t)}Ae^{Bt}
\end{equation}

where $\alpha(t)=1+\varepsilon+\alpha't$ is the pomeron trajectory, and $\varepsilon$, $\alpha'$, $A$, and $B$ are obtained from fitting the data. Recent fits include sub-leading contributions (commonly denoted as ``Reggeon'' exchange), which became pronounced at high $\xi$, but the H1 data weakly constrain them.

\begin{figure}[H]
\centering
\includegraphics[width=0.45\textwidth]{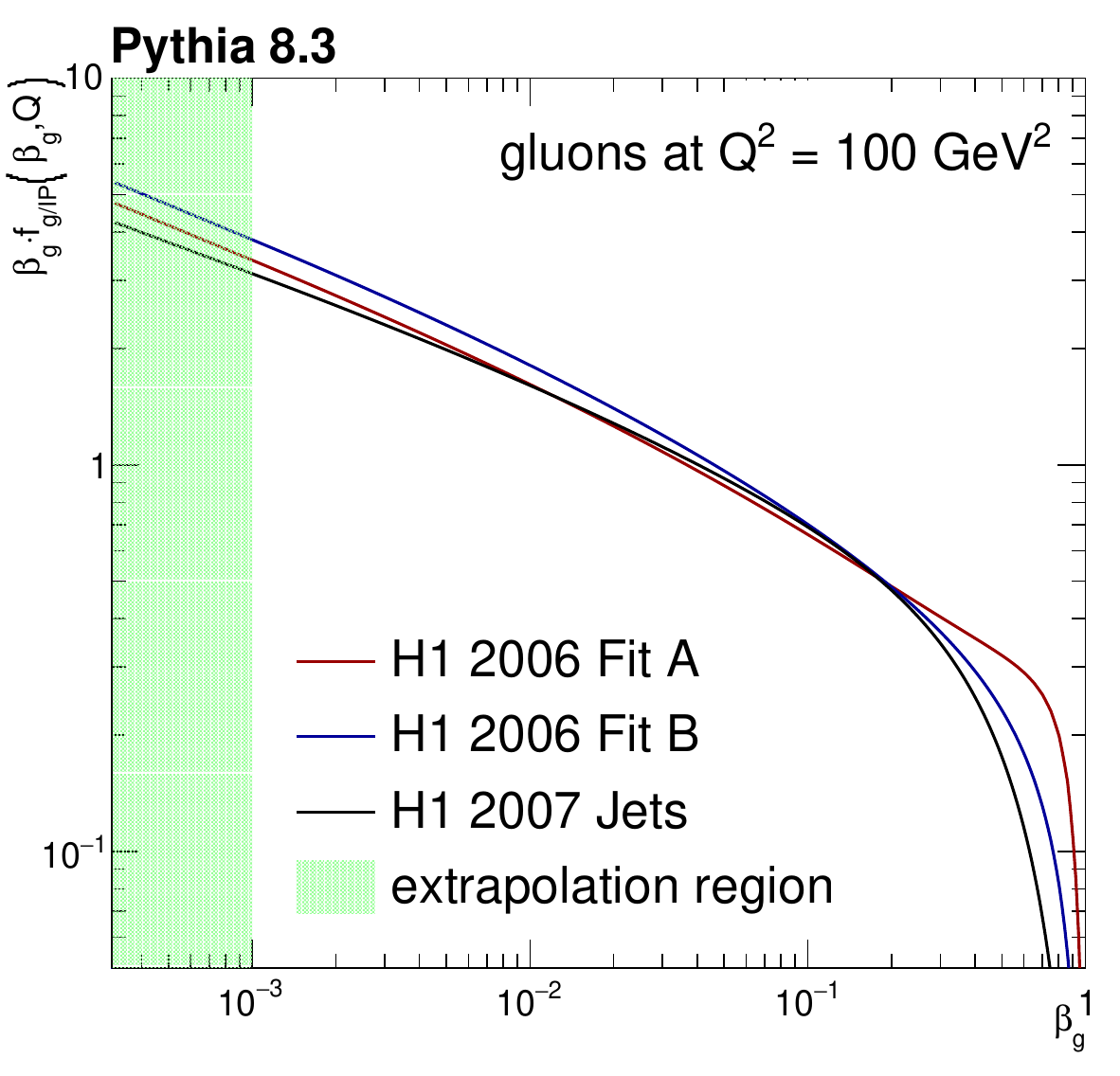}
\includegraphics[width=0.45\textwidth]{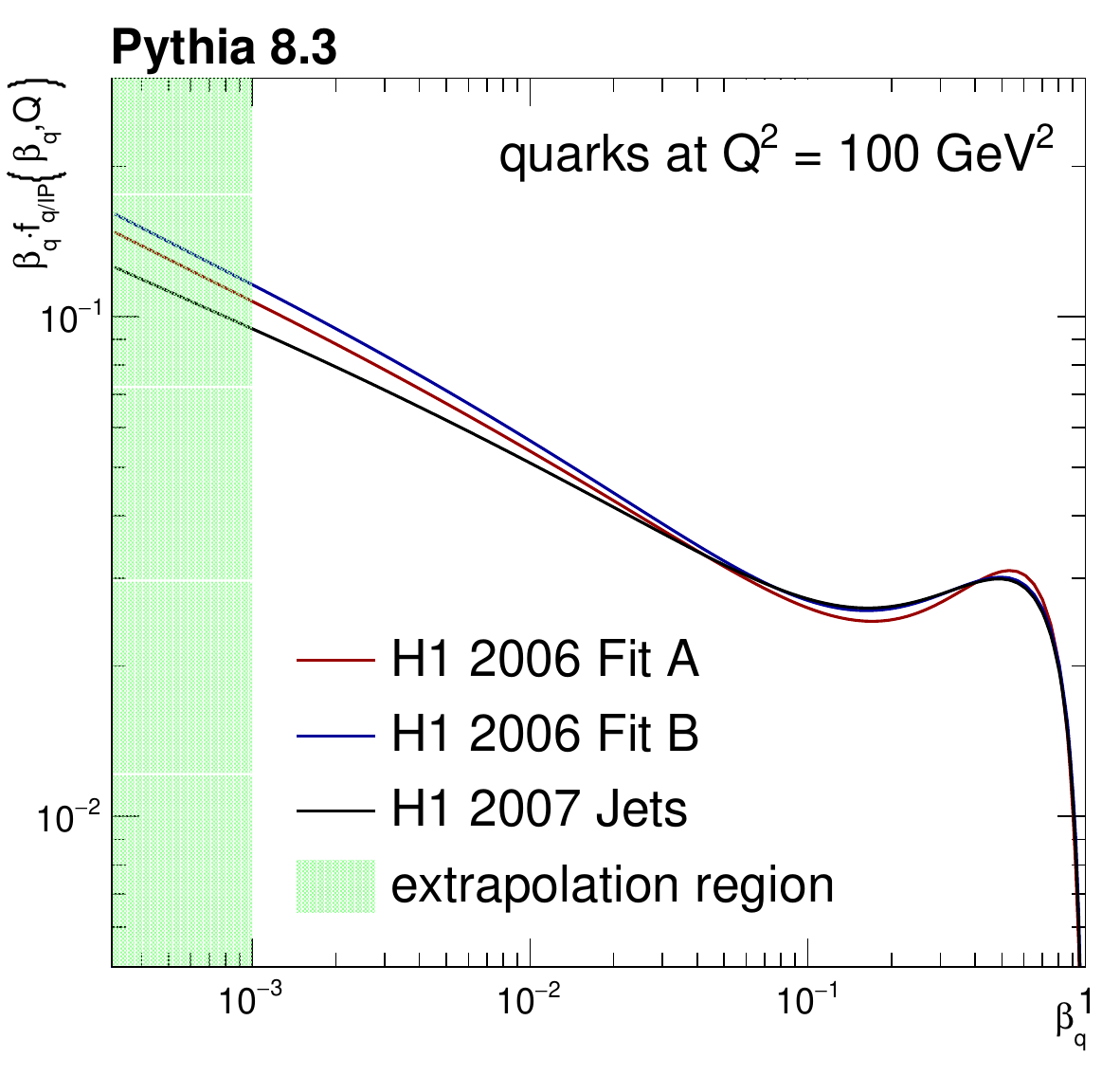}
\caption{Diffractive parton densities ($\beta\cdot f(\beta,\mu=Q)$ ) in the pomeron as a function of the fraction of the momentum carried by the gluon (\textbf{left}) or quark (\textbf{right}), extracted from a HERA fit to combined structure--function data from H1~\cite{H1:2006zyl}.}
     \label{fig:dPDF}
\end{figure}

The pomeron structure function is the probability of extracting a parton from a pomeron. In the simplest model, the most general form of the dPDF is~\cite{H1:2006zyl} 
\begin{equation}
\beta f_{i/\rm{I\!P}}(\beta)=A_i\beta^{B_i}\left(1-\beta\right)^{C_i}
\end{equation}
where $A_i$, $B_i$, and $C_i$ are fit parameters. In the latest fit to the diffractive data, the dPDFs are modeled by incorporating a light flavor quark distribution (assuming zero intrinsic densities for c- and b-quarks within the pomeron) and the gluon distribution. Figure~\ref{fig:dPDF} illustrates a few existing pomeron structure functions fitted by the H1 collaboration.  

The following MC event generators can simulate processes involving the diffractive and photon-induced production of top quark.

The SuperChic event generator~\cite{Harland-Lang:2015cta,Harland-Lang:2018iur} is used for studying central exclusive and semi-exclusive production processes in proton--proton, proton--ion, and ion--ion collisions. It implements the central exclusive production via Pomeron exchange using the improved perturbative QCD estimates provided by the ``Durham model''~\cite{Khoze:2000cy}. In the latest version, SuperChic v4, the photon-induced production of top quark pairs has also been made available~\cite{Harland-Lang:2020veo}. For photon-induced processes, the event generator applies form-factors based on the ``structure--function'' approach~\cite{Han:1992hr}.

The Madgraph5\_aMC@NLO event generator~\cite{Alwall:2014hca} is designed to generate matrix elements for both SM and beyond SM processes, including NLO QCD and EW corrections for parton-initiated processes~\cite{Hirschi:2015iia,Frederix:2018nkq}. While it can generate processes initiated by photons, it does not cover those initiated by pomerons. Elastic photon fluxes are obtained using the equivalent photon approximation (EPA)~\cite{BUDNEV1975181}, and inelastic photon-initiated collisions can be generated from $\gamma$ distribution functions inside the proton, such as, e.g., the LUXqed \mbox{one~\cite{Manohar:2017eqh}}. A novel event generator, gamma-UPC~\cite{Shao:2022cly}, has been recently introduced, which derives photon fluxes from electric dipole and charge form-factors for protons and ions and includes realistic hadronic survival probabilities for them. This model has been successfully interfaced with Madgraph5\_aMC@NLO v3. In addition, within this framework, the \mbox{$\gamma\gamma\to t\bar{t}$} process can be computed at NLO perturbative QCD accuracy.

Forward Physics Monte Carlo (FPMC)~\cite{Boonekamp:2011ky} is a specialized  Monte Carlo event generator developed for simulating exclusive and diffractive production processes. It possesses the capability to simulate all elementary $2\to 2$ and $2\to 1$ processes available in \mbox{HERWIG6.5~\cite{Corcella:2002jc}}, particularly the SM top quark pair production, and t-channel single top production. In FPMC, the original HERWIG code, which simulates two-photon exchange in electron--positron collisions, was modified. Pomeron fluxes were introduced with a particular proton structure in diffractive events (based on H1 fits) and are used in hadron collisions in this case. Recently, the anomalous exclusive production of $\gamma\gamma\to t\bar{t}$ was implemented in FPMC as well~\cite{Baldenegro:2022kaa}.

Pythia 8~\cite{Bierlich:2022pfr}, one of the most used event generators, is widely used for simulating events involving various interactions and particles, including hard diffraction~\cite{Rasmussen:2015iid}, resolved and direct photons~\cite{Helenius:2017aqz}, and photoproduction~\cite{Helenius:2019gbd}. The survival factors for Pomeron-induced processes are implemented as a part of the multiparton interaction (MPI) framework.

Table~\ref{tab1} {summarizes different MC generators and outlines the processes that can be simulated by each.} 

\begin{table}[H] 
\caption{A list of various MC generators and available top quark production modes of processes with top quarks at the final state. In the table, $X$ represent proton or pomeron remnants in a \mbox{dissociative process.}\label{tab1}}
\scriptsize
\begin{tabular}{lccccc}
\hline
\textbf{Generator}	& \boldmath{$\gamma\gamma\to t\bar{t}$}	&\boldmath{$I\!PI\!P\to t\bar{t}$} & \boldmath{$I\!PI\!P\to t\bar{t}X$}  & \boldmath{$\gamma p \to t\bar{t}X$} & \boldmath{$I\!P p \to t\bar{t}X$} \\
\hline
Superchic~v4		& \checkmark & \checkmark & ---  & ---			& --- \\
MadGraph/gamma-UPC	& \checkmark & --- & --- & \checkmark & --- \\
FPMC		& \checkmark & \checkmark	& \checkmark & \checkmark		& \checkmark \\
Pythia8	& \checkmark & ---	& --- & \checkmark		& \checkmark \\
\hline

& \boldmath{$I\!PI\!P\to tqX$}	&\boldmath{$I\!PI\!P\to tWX$} & \boldmath{$\gamma p \to tqX$} & \boldmath{$\gamma p \to tWX$} & \boldmath{$I\!P p \to tqX$} \\
\hline
MadGraph/gamma-UPC 	& --- & --- & \checkmark & \checkmark & --- \\
FPMC		& \checkmark & --- & \checkmark  & --- & \checkmark \\
Pythia8		& --- & --- & \checkmark  & --- & \checkmark \\

\hline
\end{tabular}
\end{table}


\section{Tagging diffractive and photon-induced processes}

\begin{figure}[H]
\includegraphics[width=0.95\textwidth]{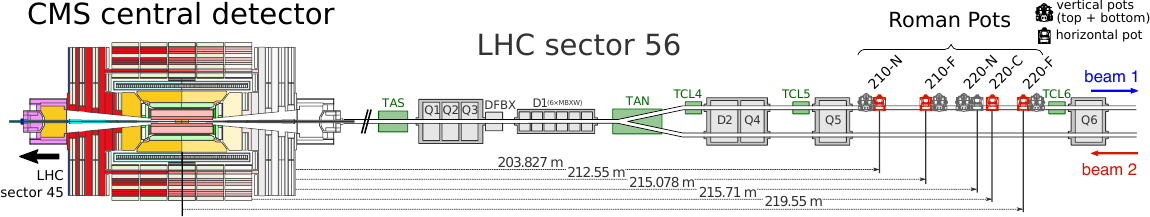}
\centering
\caption{Schematic layout of the CMS Proton Precision Spectrometer (PPS) located in LHC sector 56. The accelerator magnets are displayed in gray, while the collimator system parts are highlighted in green. The detector units, marked in red, are those used by PPS during Run 2. The figure is taken from Ref.~\cite{CMS:2023gfb}.\label{fig:pps-layout}}\end{figure}

Diffractive and photon-initiated processes often exhibit low hadronic activity, and some events are characterized by an intact proton emerging from the primary interaction. Forward Proton Detectors (FPD) are utilized at the LHC to identify these events by detecting forward protons. These detectors, such as the ATLAS Forward Proton detector (AFP)~\cite{Adamczyk:2015cjy} and the CMS-TOTEM Precision Proton Spectrometer (CT-PPS)~\cite{CMS:2014sdw}, are positioned approximately 200 m from the proton--proton interaction point. In diffractive or photon-initiated interactions, protons lose a fraction of their nominal momentum and are deflected differently by the LHC magnets, diverting them from the main bunch of protons. The FPD are near-beam detectors usually housed in Roman Pots vessels~\cite{Amaldi:1972uw} which could approach the proton beam up to a few mm, aiming to measure the slight displacement of protons that participated in the interaction. Figure~\ref{fig:pps-layout} illustrates a schematic layout of the beamline between the interaction point and the FPDs installed in LHC sector 56, corresponding to the negative $z$ direction in the CMS.

The FPDs are equipped with tracking and timing capabilities that determine the proton scattering angle and momentum loss based on the hit position of protons in the detector planes. The hit position of a proton is affected by its momentum loss, denoted by $\xi = \Delta p_z/p$, two scattered angles at the interaction point $\left( \theta_x^*, \theta_y^*\right)$, and coordinates of the proton--proton collision vertex situated on the plane perpendicular to the beam trajectory, denoted by $\left(x^*, y^*\right)$:

\begin{equation}
\begin{aligned}
\delta x &= x_D(\xi)+v_x(\xi)\cdot x^*+L_x(\xi)\cdot \theta_x^*  \\
\delta y &= y_D(\xi)+v_y(\xi)\cdot y^*+L_y(\xi)\cdot \theta_y^* 
\end{aligned}
\label{eq:proton_transport}
\end{equation}
where $D_x$ and $D_y$ are the horizontal and vertical dispersion, respectively. The terms $v_x$ and $v_y$ stand for the horizontal and vertical magnifications, respectively, while $L_{x,y}$ are the effective lengths. These parameters are functions of the proton's momentum loss at different positions from the interaction point and are determined by simulating the proton's trajectory in the LHC magnetic field, as described in reference~\cite{TOTEM:2022vox}. 

A minimum of five independent spatial measurements of the scattered proton are required to accurately determine proton kinematics, necessitating at least three tracking stations. However, even with just two tracking stations, it is possible to reconstruct proton kinematics by approximating Equation \eqref{eq:proton_transport}, such as considering $x^*=0$, as demonstrated in reference~\cite{TOTEM:2022vox}. 

The timing detectors measure the arrival times of the protons to the FPD. The time difference between the arrival of two protons from the same interaction vertex, in double-tagged events, is tied to the z-position of their production vertex. The correlation between the vertex position deduced from the proton arrival times and the vertex position determined from final state particles produced in association with the protons serves as an effective tool to separate diffractive interactions from non-diffractive ones~\cite{Cerny:2020rvp}. This is particularly useful in rejecting protons originating from additional proton--proton interactions occurring during the same bunch crossing in high-intensity runs (high pileup runs) at the LHC. In addition to double-tagged events, the proton timing detectors can differentiate between pileup protons in single-tagged events, provided that the arrival time of the final state particles is also measurable~\cite{Pasechnik:2023mdd}.

\section{Diffractive and photo-production of top quarks}

\subsection{Single-diffraction and photoproduction of top quark pairs}
The diffractive production of top quarks can constitute up to a few percent of the total inclusive production cross-section. This aspect is crucial when aiming for precision measurements involving top quarks or when searching for new physics phenomena. For a top quark mass of $m_\text{top}=172.5$~GeV, the inclusive production cross-section of top quark pairs, calculated at next-to-next leading order (NNLO) in QCD with resummation at next-to-next-to-leading logarithmic (NNLL) soft gluon terms~\cite{Czakon:2011xx}, and more recently with third-order soft gluon corrections and the additional inclusion of electroweak corrections~\cite{Kidonakis:2023juy}, ranges from around 180 to around 1000~pb, for proton--proton collision energies of $\sqrt{s}=7$~TeV and $\sqrt{s}=14$~TeV, respectively. The dominant diffractive production modes of top quark pairs are via single-diffractive dissociation and photoproduction, with their respective leading order (LO) cross-sections at $\sqrt{s}=13$~TeV calculated to be 5~pb and 1.45~pb, respectively~\cite{Howarth:2020uaa}. 

\begin{figure}[H]
\centering
\includegraphics[width=0.45\textwidth]{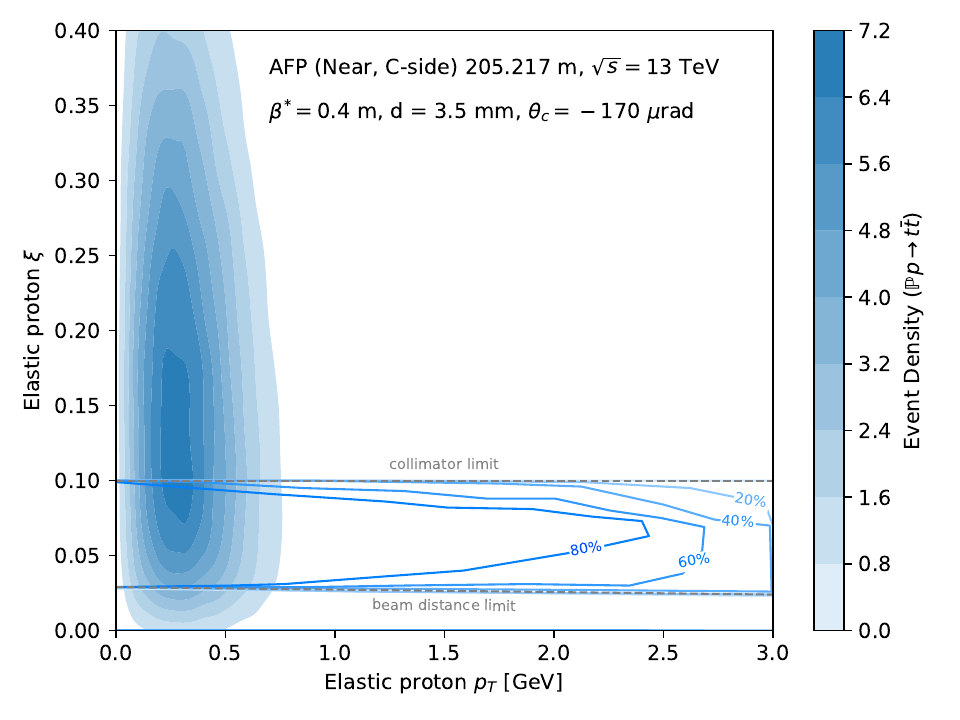}
\includegraphics[width=0.45\textwidth]{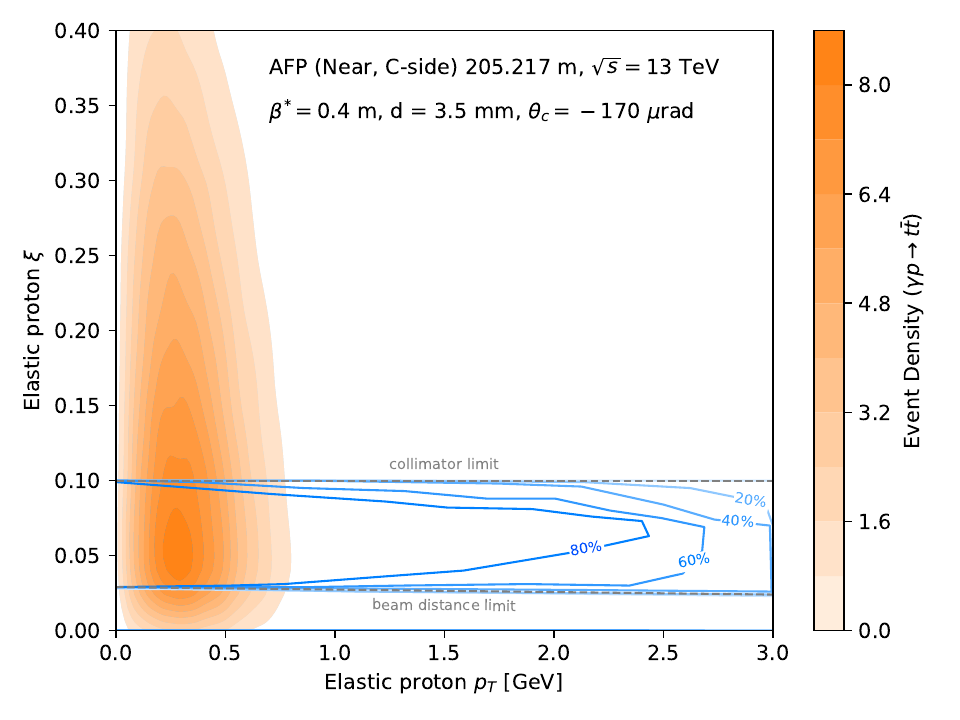}
\caption{The event density for the production of top quark pairs in single-diffractive (left) and photoproduction (right) modes as a function of the proton's $p_T$ and relative momentum loss ($\xi$). For context, acceptance contours for ATLAS AFP detectors using LHC Run 2 optics are superimposed for comparison. Figures are taken from reference~\cite{Howarth:2020uaa}. \label{fig:proton_kinematics_SD}}
\end{figure}

The study outlined in reference~\cite{Howarth:2020uaa} investigated the single diffractive and photoproduction processes. The analysis assumed an FPD acceptance to an intact proton with momentum loss between 3\% and 10\%. This span is defined by a set of constraints; the lower limit is determined by the minimum distance of the detectors from the beam, while the upper limit is constrained by the beam collimators that shield the magnets from the intense radiation. The study revealed that, for photon-induced and single-diffractive processes, the acceptance rates are 30\% and 20\%, respectively, driven by the different photon and pomeron density fluxes. The intact proton kinematics for the pomeron- and photon-initiated production of top quark pairs were computed for the proton--proton beam conditions used during LHC Run 2 (2015--2018) and are depicted in Figure~\ref{fig:proton_kinematics_SD}.

Based on the $t\bar{t}$ selection criteria and reconstruction efficiencies obtained in the measurement of differential cross sections of top quark pair production, in association with jets in pp collisions at $\sqrt{s}=13$ by the ATLAS collaboration~\cite{ATLAS:2018acq}, event yields were calculated to be 150 $\pm$ 20 for single diffractive and 94 $\pm$ 3 photo-production of top quark pairs, given an integrated luminosity of 1 fb~$^{-1}$. At the LHC, the predominant background for diffractive and photon-induced events stems from the multiple proton--proton collisions occurring in a single bunch crossing. This results in a combinatorial background, where a non-diffractive collision may be accompanied by additional diffractively scattered protons from other soft proton--proton interactions. To observe top quarks produced in association with an intact proton, a dataset with a low average number of interactions per bunch crossing is desired, ideally below 0.01. However, no data samples meeting these conditions have been collected so far.

\subsection{Single-diffraction production of single top quark}

\begin{figure}[H]
\centering
\includegraphics[width=0.43\textwidth]{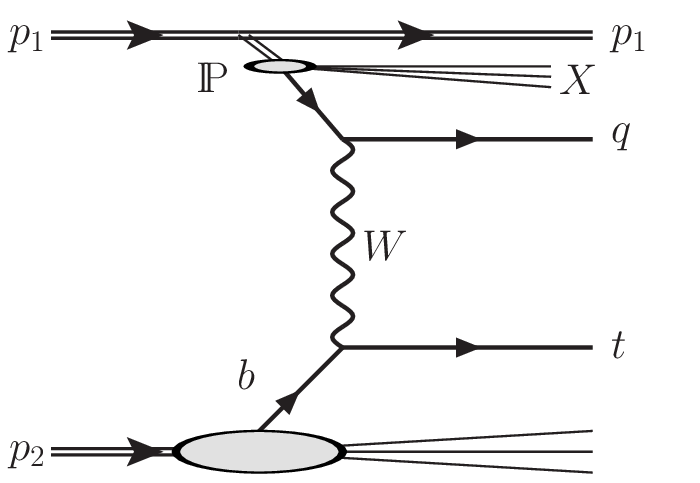}
\includegraphics[width=0.43\textwidth]{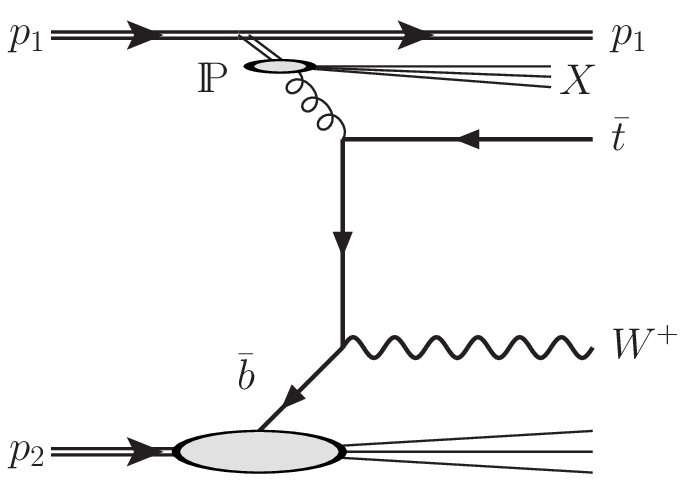}
\caption{Representative diagrams of the single diffractive production of t-channel single top (\textbf{left}), and the W-associated production of single top (\textbf{right}).\label{fig:ST_diagrams}}
\end{figure}

The study of the production of single top quarks at the LHC holds significant importance, as it probes the heavy quark content of the proton (proton PDF). This stems from the fact that the process involves interactions of heavy quarks originating from dissociated protons. The diffractive production of single top quarks offers further insights into the heavy quark content of the pomeron. Figure~\ref{fig:ST_diagrams} illustrates the diagrams representing \mbox{these interactions.}

\begin{figure}[H]
\includegraphics[width=0.5\textwidth]{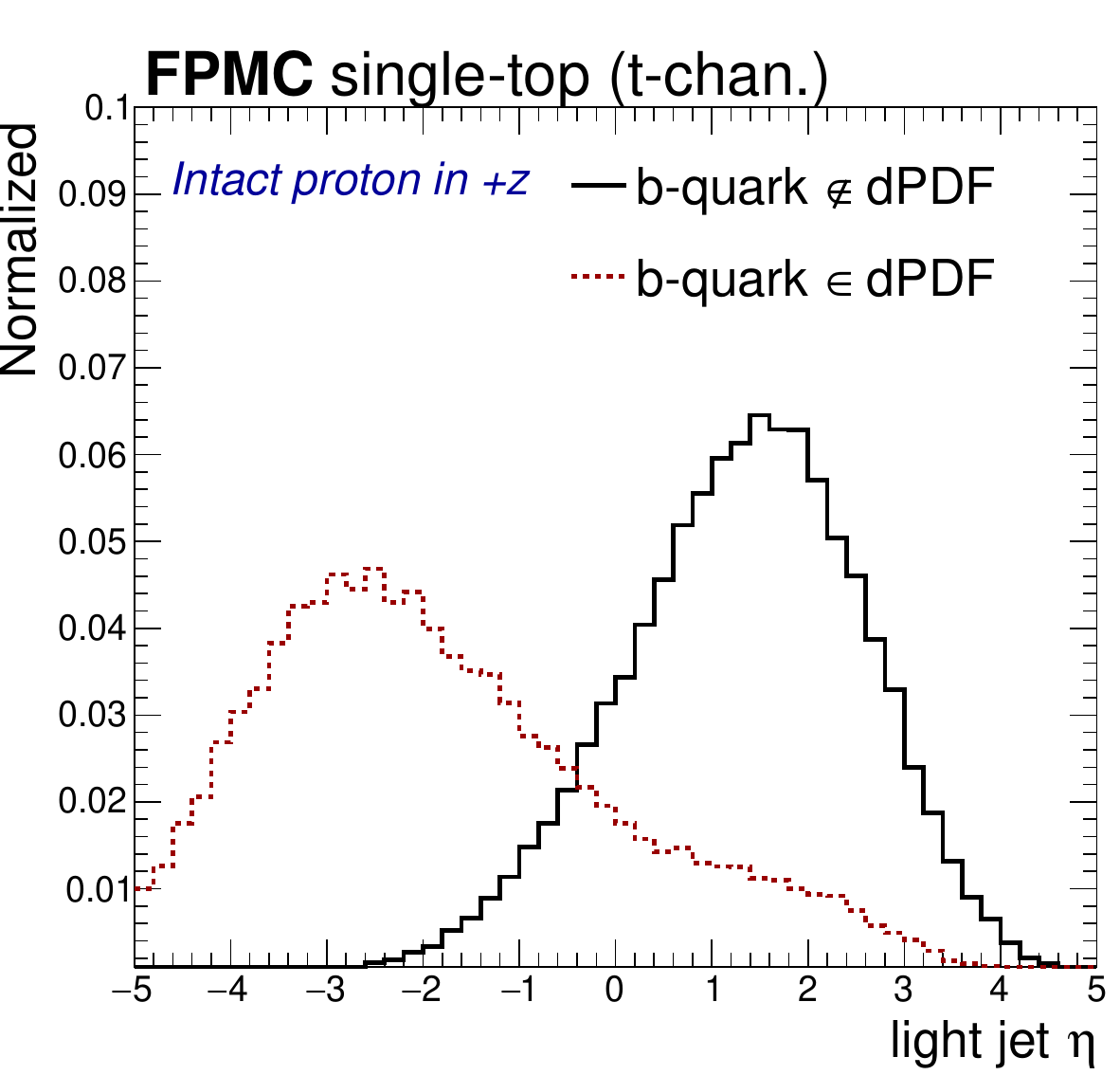}
\centering
\caption{Light jet pseudorapidity in hard single diffractive $pp\to tq$ production process, distinguishing between different intrinsic quark flavors of the pomeron, either b-quarks (dashed red line) or gluons (solid black line).\label{fig:xi_ST}}
\end{figure}

Besides its use as a valuable tool for probing PDFs, the single diffractive production of single-top quarks can also shed light on the intrinsic pomeron content of heavy flavor quarks. The origin of the heavy flavor quark, whether from the proton or the pomeron, can be discerned by measuring the kinematics of the light jet in the \textit{t}-channel single-top quark production. The difference between diffractive and non-diffractive Parton Distribution Functions (PDFs) leads to a disparity in the hardness of partons originating from non-diffractive protons. As a result, the central system is typically boosted in the direction of the non-diffractive proton when partons from the latter are more energetic. Therefore, by tagging the light jet and the intact proton, one can determine whether it is a proton or a pomeron with intrinsic heavy flavor content. The pseudorapidity distribution of the light jet for generated single-diffractive single-top-quark events, where the bottom quark is initiated by a diffractive or non-diffractive proton, is shown in Figure~\ref{fig:xi_ST}.

Given the low cross-section of the single diffractive production of single-top events and the high density of gluons within the pomeron, such processes have not received significant attention in the literature until the present day.

\subsection{Double pomeron exchange processes}

\begin{figure}[H]
\centering
\includegraphics[width=.31\textwidth]{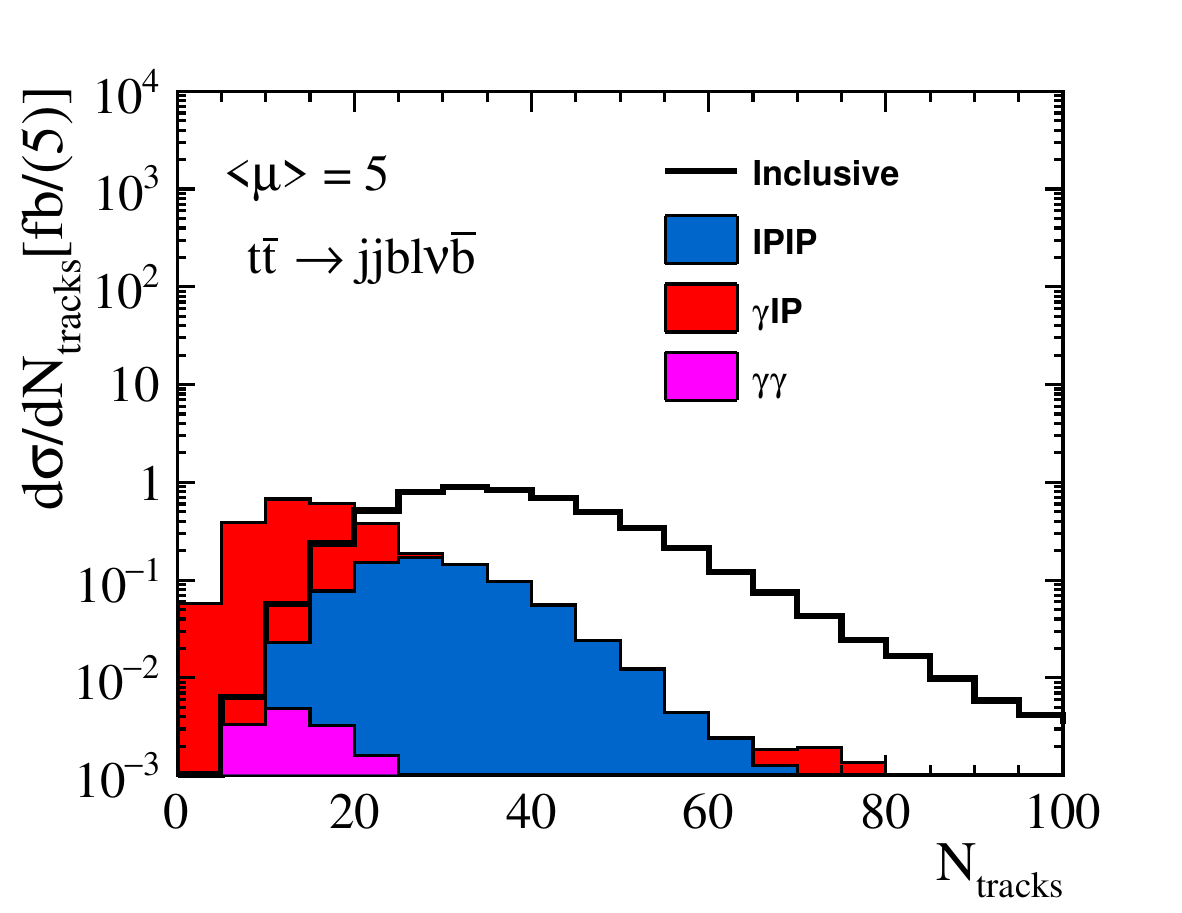}
\includegraphics[width=.31\textwidth]{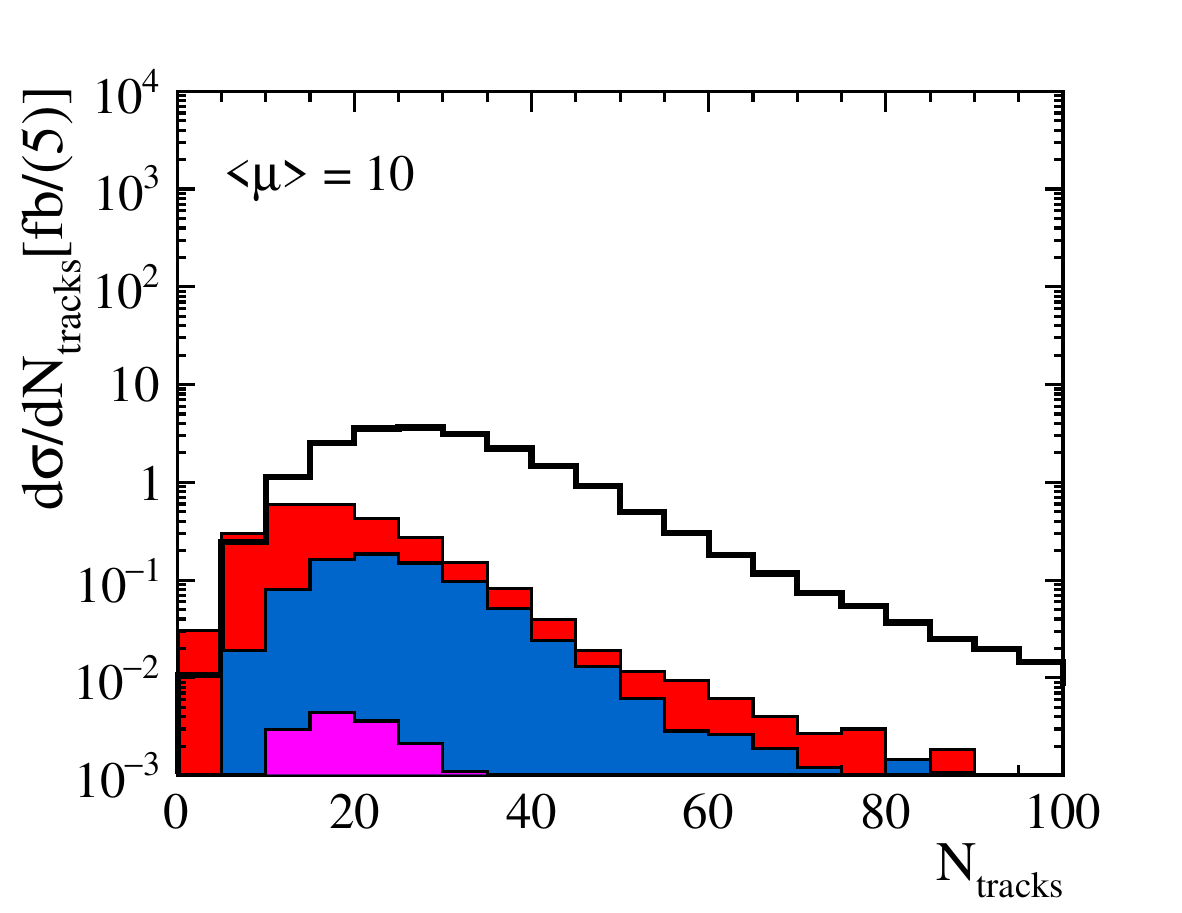}
\includegraphics[width=.31\textwidth]{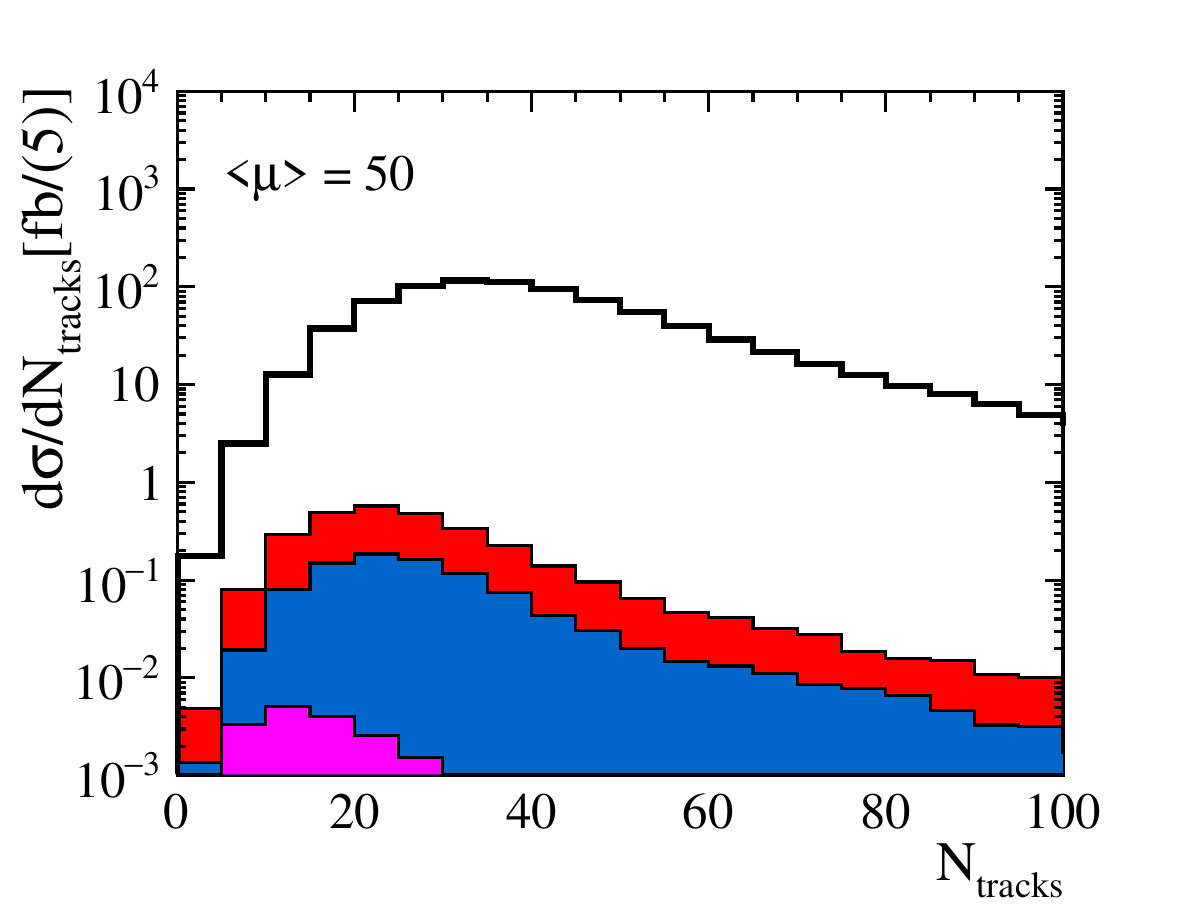}
\caption{Distribution of the number of tracks for three different pileup profiles after event preselection, as detailed in the referenced publication. The plot is taken from reference~\cite{Goncalves:2020saa}.\label{fig:ntracks}}
\end{figure}

The diffractive events associated with two intact protons can arise when two color-singlets are exchanged. These events typically have a lower combinatorial background, but the signal cross-section drops substantially. As discussed in references~\cite{Goncalves:2020saa,Martins:2022dfg}, the production cross-section for $\gamma\gamma\to t\bar{t}+X$, $\gamma I\!P\to t\bar{t}+X$, and $I\!PI\!P\to t\bar{t}+X$ are 0.34~fb, 52~fb, and 28.4~fb, respectively.  The authors of reference~\cite{Goncalves:2020saa} estimated the sensitivity to diffractive production of top quarks in so-called semi-leptonic $t\bar{t}$ decays, $t\bar{t}\to jjb\ell\nu_\ell\bar{b}$, where one top quark decays hadronically into two light quarks and a b-quark, and the other into a b-quark with a W boson, which then decays leptonically into a lepton and neutrino. The size of this combinatorial background depends exponentially on the number of pileup collisions. Ab analysis followed the semi-leptonic event selection: four jets with $p_T>$25~GeV and $|\eta|<2.5$, one lepton (electron or muon) with $p_T>$25~GeV and $|\eta|<2.5$, at least two jets b-tagged jets, FPD acceptance between 1.5\% and 15\%, and a limited number of tracks associated with a primary interaction vertex ($N_\text{TRK}$). Due to the nature of the hard color-singlet exchange, the signature of diffractive interactions is in low hadronic activity, and $N_\text{TRK}$ is typically used as a discriminating variable. The distribution of the number of tracks for different pileup events per interaction is shown in Figure~\ref{fig:ntracks}. 

The measurement of the diffractive production of top quarks requires a significant number of signal events and minimal background contamination. The authors of reference~\cite{Goncalves:2020saa} have shown that the statistical significance for observing a diffractive signature was $\sigma=11$ for a pileup rate of $\mu=5$ and an integrated luminosity of 10 fb$^{-1}$. However, this significance drops to $\sigma=6$ with a pileup rate of $\mu=10$ and an integrated luminosity of 30 fb$^{-1}$. 

In a subsequent study, the authors considered a pileup scenario of 200 interactions per bunch crossing, which is relevant for the High Luminosity LHC (HL-LHC) conditions~\cite{Martins:2022dfg}. To mitigate the effects of pileup, the authors integrated proton time of flight (ToF) measurements, effectively rejecting the protons stemming from pileup interaction. With a presumed timing resolution of 10 ps, the analysis managed to attain a statistical significance of $\sigma=3$ even in this high pileup scenario, using the full HL-LHC integrated luminosity of 4000~fb$^{-1}$.

These studies emphasize the challenges posed by the pileup in measuring the diffractive production of top quarks and the importance of developing techniques to mitigate these effects in order to achieve a high sensitivity to the diffractive production of top quarks.

\subsection{Central exclusive production of top quarks}

The central exclusive production of top quark pairs has the lowest production cross-section among different diffractive and photon-induced production processes. Typical cross-sections span between 0.1 and 0.4 fb~\cite{deFavereaudeJeneret:2009db,Fayazbakhsh:2015xba,dEnterria:2009cwl,Goncalves:2020saa,Howarth:2020uaa,Luszczak:2018dfi,Shao:2022cly} for elastic photon exchange and of the order of 0.01--0.001~fb for Pomeron exchange~\cite{Howarth:2020uaa,Luszczak:2018dfi}. One of the prominent advantages of proton tagging in central exclusive production processes is that the beam kinematics are exclusively transferred to the top quarks and protons. Given the absence of other intermediary particles, there is a direct correlation between the fractional momentum loss of the scattered proton and the top quarks, which can be expressed as follows:

\begin{equation}
\xi(t,\bar{t}) = \frac{1}{\sqrt{s}} \sum_{t,\bar{t}} \left[E_t \pm p_{Z} \right], 
\label{eq:proton_correlation}
\end{equation}

where $s$ denotes the center-of-mass energy of the collision, and the two solutions for $\pm p_{Z}$ pertain to the protons moving in the positive or negative $z$ direction. Equation \eqref{eq:proton_correlation} can be reformulated in terms of di-top mass ($m_{t\bar{t}}$) and rapidity ($ y_{t\bar{t}}$), yielding the following:

\begin{equation}
m_{t\bar{t}} = \sqrt{s\xi_1\xi_2},~~~~~~~~~ y_{t\bar{t}} = \frac{1}{2}\log(\xi_1\xi_2), 
\end{equation}
where $\xi_1$ and $\xi_2$ are the momentum losses of the two measured protons.

The study presented in reference~\cite{Luszczak:2018dfi} calculated both the inclusive and exclusive production cross-sections of top quark pairs via photon fusion, arguing for a considerable reduction in signal efficiency when applying a veto on charged particles or outgoing jets. They emphasized that the inclusive production of top quarks, either via gluon fusion or quark anti-quark annihilation, has an immense contribution at high pileup rates, which limits the potential to observe the Standard Model contribution of the $\gamma\gamma\to t\bar{t}$ production process at the LHC. 

In the search for the exclusive production of top quark pairs, the CMS experiment analyzed data from 2017, corresponding to an integrated luminosity of 29.4 fb$^{-1}$, collected at a center-of-mass energy of 13 TeV~\cite{CMS:2023naq}. The analysis focused on two channels: a dileptonic channel, where both top quarks decayed leptonically ($t\to b\ell\nu$), and a semileptonic channel, where one top quark decayed leptonically and the other hadronically. For the semileptonic channel, events were triggered by the presence of an electron or a muon with transverse momentum ($p_T$) above 30~GeV or 27~GeV, respectively. In addition, while the dileptonic channel used events triggered by two lepton triggers, the semileptonic channel used events triggered by a lepton and jet trigger, with the lepton $p_T$ threshold being reduced to 30~GeV and at least one jet satisfying the momentum cut of $p_T>$~35~GeV. 

In the dileptonic channel, the final selection required the presence of at least two oppositely charged leptons, where at least one of them is required to have $p_T>$~30~GeV and $|\eta| < 2.1$, and the dilepton system, where the form is required to have an invariant mass $m_{\ell\ell} > $~20~GeV. For the events with two reconstructed leptons of the same flavor, $m_{\ell\ell}$ must be outside a 30~GeV mass window around the Z boson mass peak. In the semileptonic channel, the final selection required the presence of exactly one lepton (electron or muon), at least two jets passing the b-tagging selection criteria, and at least two jets failing the b-tagging selection criteria. The b-tagging selection criteria were based on the \textit{DeepCSV} algorithm~\cite{Sirunyan:2017ezt}.

\begin{figure}[H]
\centering
\includegraphics[width=0.45\textwidth]{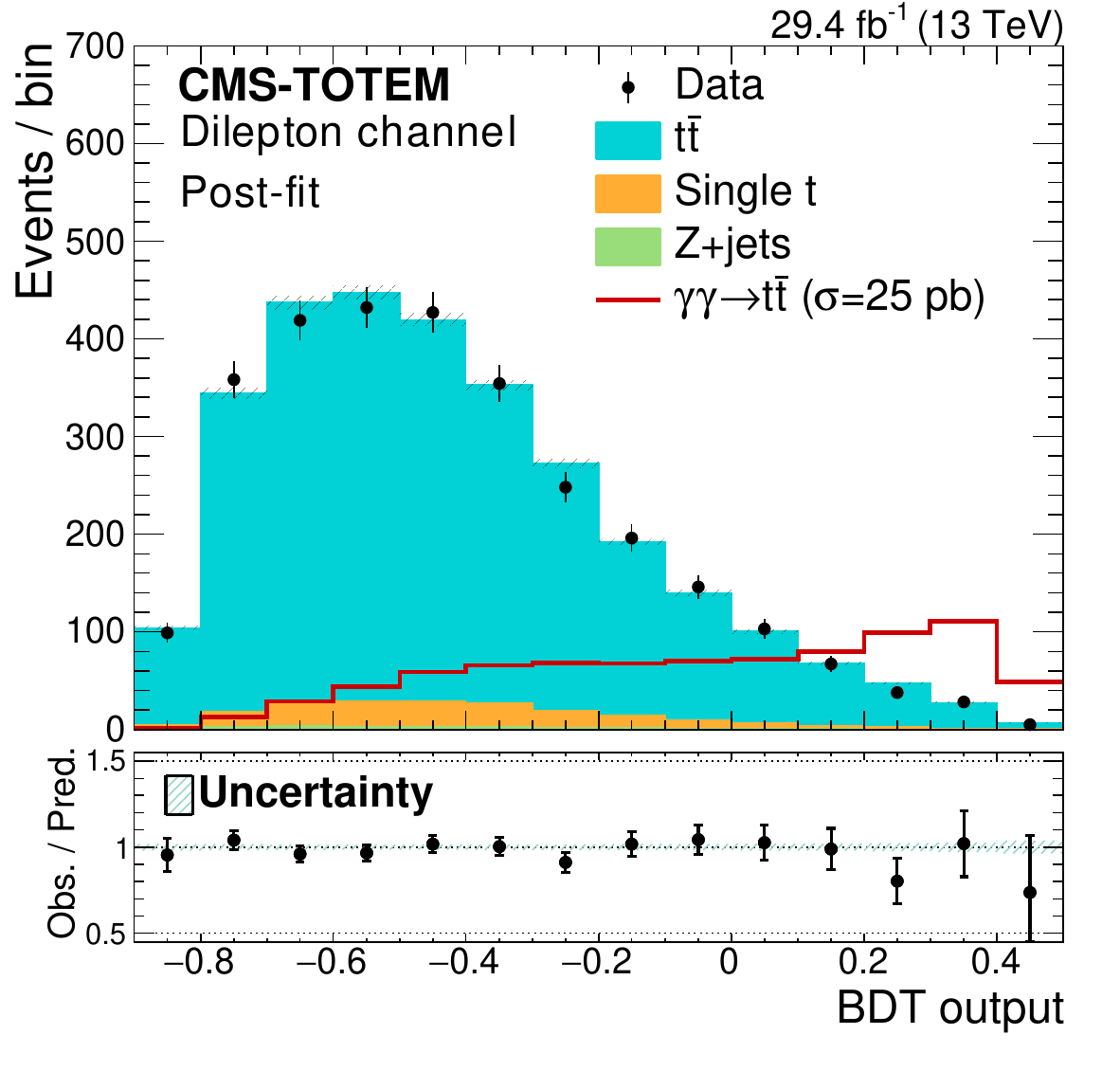}
\includegraphics[width=0.45\textwidth]{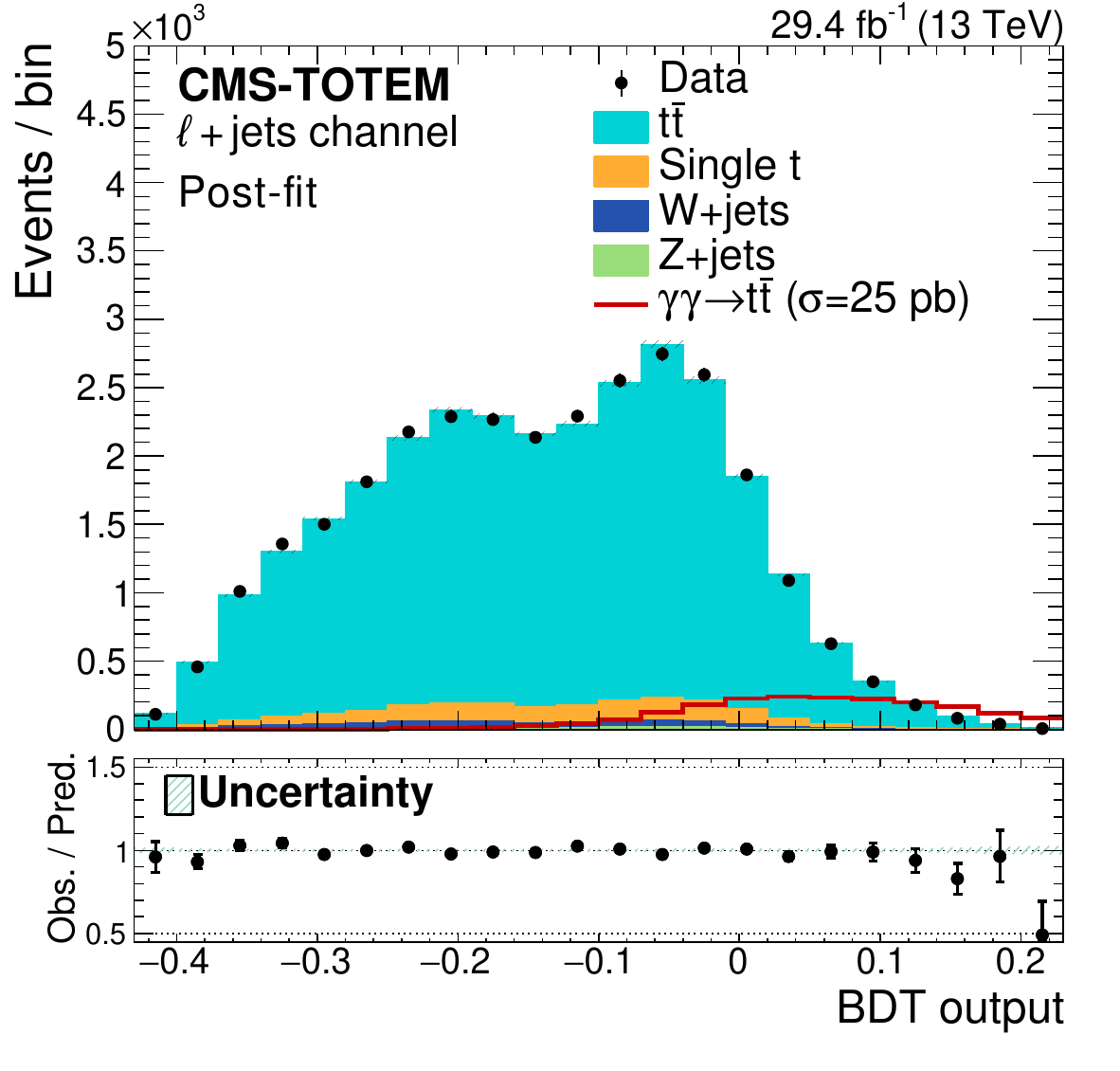}
\caption{Distribution of the BDT score in the signal region for simulated events after the fit and for data for dileptonic (\textbf{left}) and semi-leptonic (\textbf{right}) channels. The figure is taken from reference~\cite{CMS:2023naq}.\label{fig:TOP-21-007-1}}
\end{figure}

The analysis utilizes a boosted decision tree (BDT) algorithm to discriminate exclusive from inclusive production. As input variables, the kinematics of leptons and jets are used, as well as the kinematic variables obtained from proton reconstruction and the ones obtained by reconstructing the top quark pairs. The resulting BDT distributions for each channel are shown in Figure~ \ref{fig:TOP-21-007-1}.

The dominant background in the search for the exclusive production of top quark pairs is the combinatorial background, which arises from non-diffractive $t\bar{t}$ events produced in conjunction with two uncorrelated protons from pileup interactions. These events were modeled by mixing the reconstructed protons measured from the data with non-diffractive $t\bar{t}$ processes simulated in MC, normalized to the pileup proton tagging rate extracted from the data, creating a ``mixed'' MC samples with pileup proton information.

\begin{figure}[H]
\centering
\includegraphics[width=0.7\textwidth]{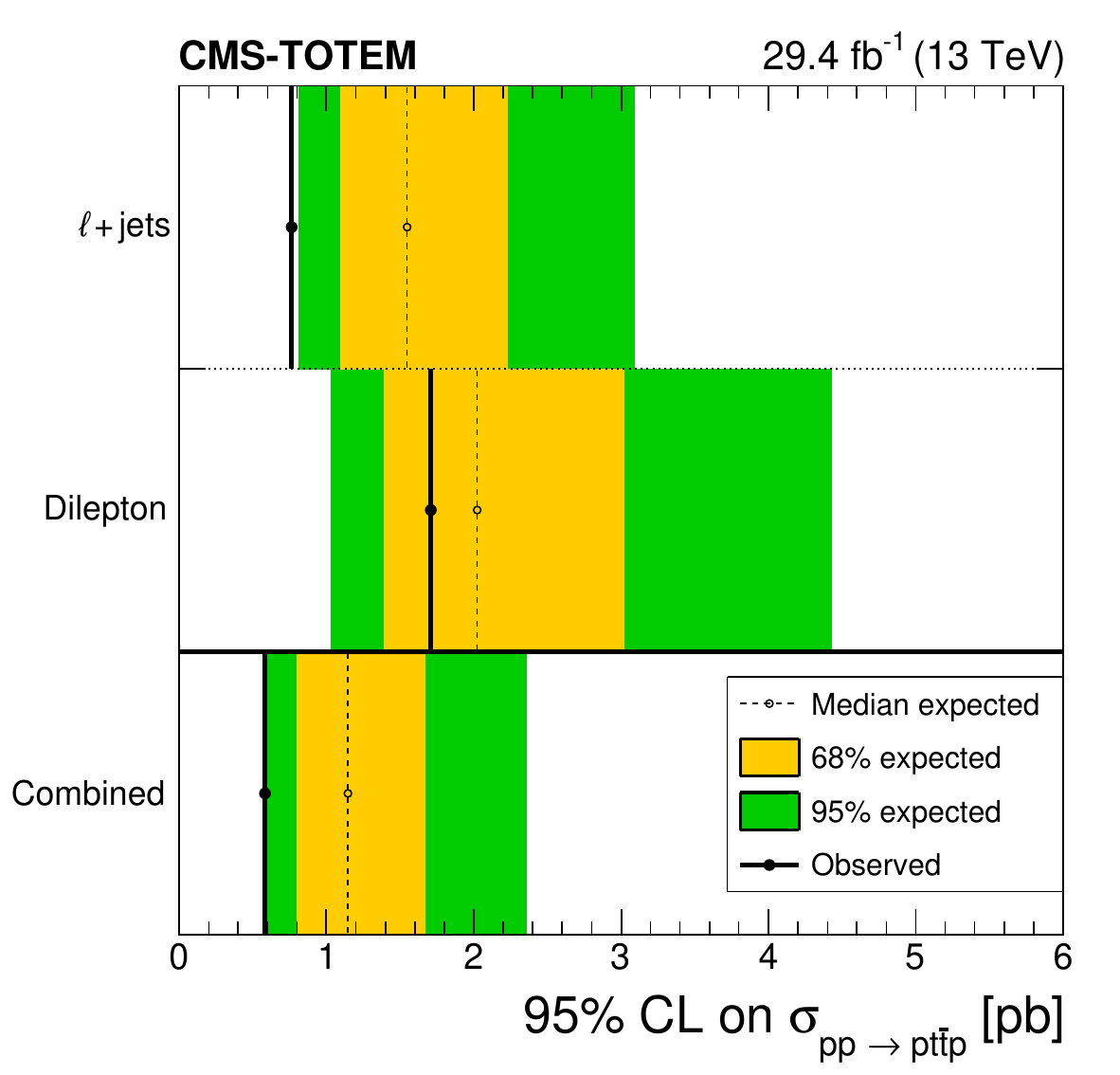}
\caption{Expected 95\% CL upper limit for the signal cross-section, for the two event selections (semi-leptonic and di-leptonic), and their combination. The green and yellow bands show the $\pm\sigma$ and $\pm 2\sigma$ intervals, respectively. The figure is taken from reference~\cite{CMS:2023naq}.\label{fig:TOP-21-007}}
\end{figure}

The probability of detecting at least one proton per arm in data ranges from 40 to 70\%, depending on the LHC magnets settings and instantaneous luminosity, results in very high background rates from non-diffractive events, as the central exclusive production of top quark pairs comprises an order of $10^{-5}$ percent of the inclusive cross-section. Due to the high combinatorial background, with an average pileup interaction rate of $\mu\sim35$, and a significantly small signal cross-section, an upper bound on the production cross-section was set at 0.59~pb at a 95\% confidence level. This corresponds to about 3000 times the SM cross-section. The resulting observed and expected limits are shown in Figure \ref{fig:TOP-21-007}. With the anticipated improvements in FPD timing capabilities and the larger amount of data expected to be collected at the HL-LHC, the potential for observing SM-exclusive top quark production will be increased~\cite{CMS:2021ncv,Pitt:2023vpo,97898112801840018}.

\section{New physics searches with photon-induced production of top quarks}

Photon-induced production of top quark offers a unique opportunity to study flavor-changing neutral currents (FCNC) within the context of the SM. In the SM, production channels such as $u\gamma\to t$ or $c\gamma\to t$ are considerably suppressed. Therefore, the observation of top quark production without any accompanying quarks or bosons might indicate the presence of photon-mediated FCNC interactions. 

Searches for single top quark production through FCNC have been conducted in reference~\cite{ATLAS:2015iqc}, focusing on gluon-initiated processes. However, the photon-induced production of direct single top quarks through FCNC was proposed and investigated in reference~\cite{Goldouzian:2016mrt}. Given the pronounced resemblance to the single top production process in the SM, a discriminator based on a neural network classifier was developed. Yet, the sensitivity was found to be comparable to the gluon-initiated processes. Nevertheless, as highlighted in reference~\cite{Howarth:2020uaa}, proton tagging in photon-initiated processes could improve the prospects for detecting FCNC interactions in photon-induced top quark production.

The central exclusive production of top quarks has a cross-section below 1~fb, necessitating a large amount of data and effective discrimination against pileup protons. Nonetheless, photon--photon fusion processes at the LHC offer opportunities to probe the SM and search for various beyond-SM physics models with anomalous $t\gamma$ couplings that could enhance production cross-section. These processes could have a distinct final state characterized by the exclusive topology, which includes the absence of proton beam remnants. Utilizing the kinematic correlation between reconstructed protons and top quarks could result in a search with low backgrounds.

The anomalous $\gamma t\bar{t}$ couplings were explored in reference~\cite{Fayazbakhsh:2015xba} in terms of its impact on the electromagnetic dipole moments of the top quarks in $pp$ collisions at the center of mass energies anticipated for HL-LHC (14 TeV) and HE-LHC (33 TeV) with integrated luminosities of 100, 300, and 3000~fb$^{-1}$. The analysis utilizes proton tagging, assuming FPD acceptance ranges from 1.5\% to either 15\% or 50\%. The finding revealed a heightened sensitivity to the electric dipole moment of the top quark, which is the source of CP violation, compared to the magnetic dipole moment. 

In reference~\cite{Baldenegro:2022kaa}, anomalous  $\gamma\gamma t\bar{t}$ couplings were examined through the lens of dimension eight operators within the framework of SM-effective field theory and also considering a new broad neutral resonance produced from the fusion of two photons that decays into a pair of top quarks, $\gamma\gamma\to\phi\to t\bar{t}$. The analysis assumed an FPD acceptance to fractional proton momentum loss between 1.5 and 20\% , and a timing resolution of 20 and 60 ps, compatible with the design scenario presented in reference~\cite{CMS:2014sdw}. Proton time of flight (ToF) measurements in two tagged proton events can be used to reconstruct vertex $z$ coordinate ($v_z$) and vertex production time ($v_t$), which can be computed from the following:

\begin{equation}
v_t = \frac{c}{2}\left(t_+ - t_-\right),~~~~~~~~~ v_t = \frac{1}{2}\left(t_+ + t_-\right) - \frac{z_\text{PPS}}{c}, 
\end{equation}

where $c=0.299792$~mm/ps denotes the speed of light, $t_+$ ($t_+$) indicates the ToF of the proton with positive (negative) $p_z$, and $z_\text{PPS}$ is the distance from the interaction point to the FPD timing plane. 

The study underscored a significant background suppression---about two orders of magnitude---achieved by incorporating proton ToF measurements with timing detectors having a nominal resolution of 20 ps. This highlights the pivotal role of FPD timing capabilities in refining the exclusive production of top quarks to search for physics beyond the SM. The sensitivity in terms of the dimension-8 operator coefficients was of the order of 10~TeV$^{-4}$. Furthermore, a scenario of a broad neutral scalar with a mass $m$ and two typical couplings to photons was considered:

\begin{equation}
f_{\gamma\gamma} = \frac{m}{4\pi},~~~~~~~~~ \Gamma_{\gamma\gamma}=4\pi m~~~~~~~~~\text{(Maximally broad width)}
\end{equation}

\begin{equation}
f_{\gamma\gamma} = \frac{m}{\sqrt{4\pi}},~~~~~~~~~ \Gamma_{\gamma\gamma}= m~~~~~~~~~\text{(Moderately broad width)}
\end{equation}

\begin{figure}[H]
\centering
\includegraphics[width=0.5\textwidth]{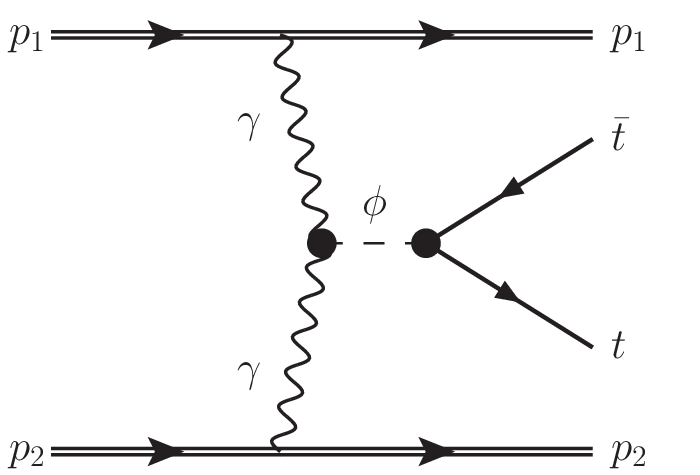}
\caption{Feynman diagram for the exclusive production of $\phi$ via photon fusion; it decays to a pair of top quarks.\label{fig:yy_phi_tt}}
\end{figure}

The neutral scalar can be considered broad for both scenarios since $\Gamma>m$. A Feynman diagram for the exclusive production of new particle $\phi$ with the decay to a pair of top quarks is shown in Figure~\ref{fig:yy_phi_tt}.

\begin{figure}[H]
\centering
\includegraphics[width=0.85\textwidth]{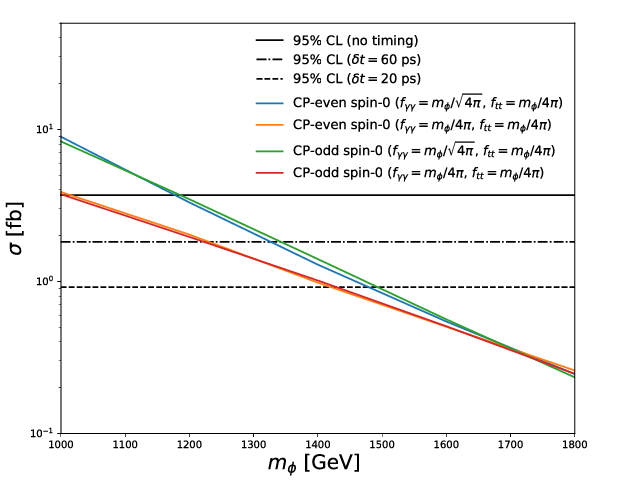}
\caption{Projected sensitivity to the $pp\to p~t\bar{t}~p$ cross-section at 95\% CL as a function of the mass of the neutral scalar. The figure is taken from reference~\cite{Baldenegro:2022kaa}. \label{fig:phi_sens}}
\end{figure}

In this scenario as well, the timing detectors played a crucial role in amplifying the search sensitivity to resonance mass of $m_\phi>$~1.5~TeV. The projected sensitivity to new scalar $\phi$ in the exclusive $t\bar{t}$ analysis is shown in Figure~\ref{fig:phi_sens}.

\section{Conclusions}

The production of top quarks at the LHC is a primary area of interest in contemporary particle physics. Most analyses primarily focus on production modes initiated by quarks or gluons. However, there is a relatively uncharted territory concerning production modes initiated by color-neutral objects like pomerons or photons. A deep understanding of these processes is vital for precision-driven studies of the top quark's properties. Furthermore, several photon-induced production modes involving top quarks at the final state could broaden the phase space explored in searches for physics beyond the SM, offering substantial discovery potential at the LHC and beyond.


\begin{thebibliography}{999}

\bibitem{Donnachie:2002en}
S.~Donnachie, H.~G. Dosch, O.~Nachtmann and P.~Landshoff, \emph{{Pomeron
  physics and QCD}}, vol.~19. Cambridge University Press, 12, 2004.

\bibitem{Collins_1984}
P.~D.~B. Collins and T.~P. Spiller, \emph{A model for diffractive top-quark
  production}, \href{https://doi.org/10.1088/0305-4616/10/12/006}{\emph{Journal
  of Physics G: Nuclear Physics} {\bfseries 10} (1984) 1667}.

\bibitem{Collins:1989gx}
J.~C. Collins, D.~E. Soper and G.~F. Sterman, \emph{{Factorization of Hard
  Processes in QCD}},
  \href{https://doi.org/10.1142/9789814503266_0001}{\emph{Adv. Ser. Direct.
  High Energy Phys.} {\bfseries 5} (1989) 1}
  [\href{https://arxiv.org/abs/hep-ph/0409313}{{\ttfamily hep-ph/0409313}}].

\bibitem{Ingelman:1984ns}
G.~Ingelman and P.~E. Schlein, \emph{{Jet Structure in High Mass Diffractive
  Scattering}}, \href{https://doi.org/10.1016/0370-2693(85)91181-5}{\emph{Phys.
  Lett. B} {\bfseries 152} (1985) 256}.

\bibitem{donnachie1988hard}
A.~Donnachie and P.~V. Landshoff, \emph{Hard diffraction: production of high pt
  jets, w or z, and drell-yan pairs}, \href{https://doi.org/10.1016/0550-3213(88)90423-3}{\emph{Nuclear Physics B} {\bfseries 303}
  (1988) 634}.

\bibitem{H1:2006zyl}
{\scshape H1} collaboration, \emph{{Measurement and QCD analysis of the
  diffractive deep-inelastic scattering cross-section at HERA}},
  \href{https://doi.org/10.1140/epjc/s10052-006-0035-3}{\emph{Eur. Phys. J. C}
  {\bfseries 48} (2006) 715}
  [\href{https://arxiv.org/abs/hep-ex/0606004}{{\ttfamily hep-ex/0606004}}].

\bibitem{Harland-Lang:2015cta}
L.~A. Harland-Lang, V.~A. Khoze and M.~G. Ryskin, \emph{{Exclusive physics at
  the LHC with SuperChic 2}},
  \href{https://doi.org/10.1140/epjc/s10052-015-3832-8}{\emph{Eur. Phys. J. C}
  {\bfseries 76} (2016) 9} [\href{https://arxiv.org/abs/1508.02718}{{\ttfamily
  1508.02718}}].

\bibitem{Harland-Lang:2018iur}
L.~A. Harland-Lang, V.~A. Khoze and M.~G. Ryskin, \emph{{Exclusive LHC physics
  with heavy ions: SuperChic 3}},
  \href{https://doi.org/10.1140/epjc/s10052-018-6530-5}{\emph{Eur. Phys. J. C}
  {\bfseries 79} (2019) 39} [\href{https://arxiv.org/abs/1810.06567}{{\ttfamily
  1810.06567}}].

\bibitem{Khoze:2000cy}
V.~A. Khoze, A.~D. Martin and M.~G. Ryskin, \emph{{Can the Higgs be seen in
  rapidity gap events at the Tevatron or the LHC?}},
  \href{https://doi.org/10.1007/s100520000359}{\emph{Eur. Phys. J. C}
  {\bfseries 14} (2000) 525}
  [\href{https://arxiv.org/abs/hep-ph/0002072}{{\ttfamily hep-ph/0002072}}].

\bibitem{Harland-Lang:2020veo}
L.~A. Harland-Lang, M.~Tasevsky, V.~A. Khoze and M.~G. Ryskin, \emph{{A new
  approach to modelling elastic and inelastic photon-initiated production at
  the LHC: SuperChic 4}},
  \href{https://doi.org/10.1140/epjc/s10052-020-08455-0}{\emph{Eur. Phys. J. C}
  {\bfseries 80} (2020) 925}
  [\href{https://arxiv.org/abs/2007.12704}{{\ttfamily 2007.12704}}].

\bibitem{Han:1992hr}
T.~Han, G.~Valencia and S.~Willenbrock, \emph{{Structure function approach to
  vector boson scattering in p p collisions}},
  \href{https://doi.org/10.1103/PhysRevLett.69.3274}{\emph{Phys. Rev. Lett.}
  {\bfseries 69} (1992) 3274}
  [\href{https://arxiv.org/abs/hep-ph/9206246}{{\ttfamily hep-ph/9206246}}].

\bibitem{Alwall:2014hca}
J.~Alwall, R.~Frederix, S.~Frixione, V.~Hirschi, F.~Maltoni, O.~Mattelaer
  et~al., \emph{{The automated computation of tree-level and next-to-leading
  order differential cross sections, and their matching to parton shower
  simulations}}, \href{https://doi.org/10.1007/JHEP07(2014)079}{\emph{JHEP}
  {\bfseries 07} (2014) 079} [\href{https://arxiv.org/abs/1405.0301}{{\ttfamily
  1405.0301}}].

\bibitem{Hirschi:2015iia}
V.~Hirschi and O.~Mattelaer, \emph{{Automated event generation for loop-induced
  processes}}, \href{https://doi.org/10.1007/JHEP10(2015)146}{\emph{JHEP}
  {\bfseries 10} (2015) 146}
  [\href{https://arxiv.org/abs/1507.00020}{{\ttfamily 1507.00020}}].

\bibitem{Frederix:2018nkq}
R.~Frederix, S.~Frixione, V.~Hirschi, D.~Pagani, H.~S. Shao and M.~Zaro,
  \emph{{The automation of next-to-leading order electroweak calculations}},
  \href{https://doi.org/10.1007/JHEP11(2021)085}{\emph{JHEP} {\bfseries 07}
  (2018) 185} [\href{https://arxiv.org/abs/1804.10017}{{\ttfamily
  1804.10017}}].

\bibitem{BUDNEV1975181}
V.~Budnev, I.~Ginzburg, G.~Meledin and V.~Serbo, \emph{The two-photon particle
  production mechanism. physical problems. applications. equivalent photon
  approximation},
  \href{https://doi.org/https://doi.org/10.1016/0370-1573(75)90009-5}{\emph{Physics
  Reports} {\bfseries 15} (1975) 181}.

\bibitem{Manohar:2017eqh}
A.~V. Manohar, P.~Nason, G.~P. Salam and G.~Zanderighi, \emph{{The Photon
  Content of the Proton}},
  \href{https://doi.org/10.1007/JHEP12(2017)046}{\emph{JHEP} {\bfseries 12}
  (2017) 046} [\href{https://arxiv.org/abs/1708.01256}{{\ttfamily
  1708.01256}}].

\bibitem{Shao:2022cly}
H.-S. Shao and D.~d'Enterria, \emph{{gamma-UPC: automated generation of
  exclusive photon-photon processes in ultraperipheral proton and nuclear
  collisions with varying form factors}},
  \href{https://doi.org/10.1007/JHEP09(2022)248}{\emph{JHEP} {\bfseries 09}
  (2022) 248} [\href{https://arxiv.org/abs/2207.03012}{{\ttfamily
  2207.03012}}].

\bibitem{Boonekamp:2011ky}
M.~Boonekamp, A.~Dechambre, V.~Juranek, O.~Kepka, M.~Rangel, C.~Royon et~al.,
  \emph{{FPMC: A Generator for forward physics}},
  \href{https://arxiv.org/abs/1102.2531}{{\ttfamily 1102.2531}}.

\bibitem{Corcella:2002jc}
G.~Corcella, I.~G. Knowles, G.~Marchesini, S.~Moretti, K.~Odagiri,
  P.~Richardson et~al., \emph{{HERWIG 6.5 release note}},
  \href{https://arxiv.org/abs/hep-ph/0210213}{{\ttfamily hep-ph/0210213}}.

\bibitem{Baldenegro:2022kaa}
C.~Baldenegro, A.~Bellora, S.~Fichet, G.~von Gersdorff, M.~Pitt and C.~Royon,
  \emph{{Searching for anomalous top quark interactions with proton tagging and
  timing detectors at the LHC}},
  \href{https://doi.org/10.1007/JHEP08(2022)021}{\emph{JHEP} {\bfseries 08}
  (2022) 021} [\href{https://arxiv.org/abs/2205.01173}{{\ttfamily
  2205.01173}}].

\bibitem{Bierlich:2022pfr}
C.~Bierlich et~al., \emph{{A comprehensive guide to the physics and usage of
  PYTHIA 8.3}},  \href{https://arxiv.org/abs/2203.11601}{{\ttfamily
  2203.11601}}.

\bibitem{Rasmussen:2015iid}
C.~O. Rasmussen, \emph{{Hard Diffraction in Pythia 8}},
  \href{https://doi.org/10.1051/epjconf/201612002002}{\emph{EPJ Web Conf.}
  {\bfseries 120} (2016) 02002}
  [\href{https://arxiv.org/abs/1512.05872}{{\ttfamily 1512.05872}}].

\bibitem{Helenius:2017aqz}
I.~Helenius, \emph{{Photon-photon and photon-hadron processes in Pythia 8}},
  \href{https://doi.org/10.23727/CERN-Proceedings-2018-001.119}{\emph{CERN
  Proc.} {\bfseries 1} (2018) 119}
  [\href{https://arxiv.org/abs/1708.09759}{{\ttfamily 1708.09759}}].

\bibitem{Helenius:2019gbd}
I.~Helenius and C.~O. Rasmussen, \emph{{Hard diffraction in photoproduction
  with Pythia 8}},
  \href{https://doi.org/10.1140/epjc/s10052-019-6914-1}{\emph{Eur. Phys. J. C}
  {\bfseries 79} (2019) 413}
  [\href{https://arxiv.org/abs/1901.05261}{{\ttfamily 1901.05261}}].

\bibitem{Adamczyk:2015cjy}
{The ATLAS Collaboration}, \emph{{Technical Design Report for the ATLAS Forward
  Proton Detector}}, CERN-LHCC-2015-009, ATLAS-TDR-024.

\bibitem{CMS:2014sdw}
{The CMS and TOTEM Collaborations}, \emph{{CMS-TOTEM Precision Proton
  Spectrometer}}, CERN-LHCC-2014-021, TOTEM-TDR-003, CMS-TDR-13.

\bibitem{Amaldi:1972uw}
U.~Amaldi et~al., \emph{{Measurements of the proton proton total cross-sections
  by means of Coulomb scattering at the Cern intersecting storage rings}},
  \href{https://doi.org/10.1016/0370-2693(73)90277-3}{\emph{Phys. Lett. B}
  {\bfseries 43} (1973) 231}.

\bibitem{CMS:2023gfb}
{The CMS Collaboration}, \emph{{Development of the CMS detector for the CERN
  LHC Run 3}},  \href{https://arxiv.org/abs/2309.05466}{{\ttfamily
  2309.05466}}.

\bibitem{TOTEM:2022vox}
{CMS and TOTEM Collaborations}, \emph{{Proton reconstruction with the CMS-TOTEM
  Precision Proton Spectrometer}}, \href{https://doi.org/10.1088/1748-0221/18/09/P09009}{\emph{JINST} {\bfseries
  18} (2023) P09009} [\href{https://arxiv.org/abs//2210.05854}{{\ttfamily
  2210.05854}}].

\bibitem{Cerny:2020rvp}
K.~\v{C}ern\'y, T.~S\'ykora, M.~Ta\v{s}evsk\'y and R.~\v{Z}leb\v{c}\'\i{}k,
  \emph{{Performance studies of Time-of-Flight detectors at LHC}},
  \href{https://doi.org/10.1088/1748-0221/16/01/P01030}{\emph{JINST} {\bfseries
  16} (2021) P01030} [\href{https://arxiv.org/abs/2010.00237}{{\ttfamily
  2010.00237}}].

\bibitem{Pasechnik:2023mdd}
R.~Pasechnik and M.~Ta\v{s}evsk\'y, \emph{{Multi-dimensional hadron structure
  through the lens of gluon Wigner distribution}},
  \href{https://arxiv.org/abs/2310.10793}{{\ttfamily 2310.10793}}.

\bibitem{Czakon:2011xx}
M.~Czakon and A.~Mitov, \emph{{Top++: A Program for the Calculation of the
  Top-Pair Cross-Section at Hadron Colliders}},
  \href{https://doi.org/10.1016/j.cpc.2014.06.021}{\emph{Comput. Phys. Commun.}
  {\bfseries 185} (2014) 2930}
  [\href{https://arxiv.org/abs/1112.5675}{{\ttfamily 1112.5675}}].

\bibitem{Kidonakis:2023juy}
N.~Kidonakis, M.~Guzzi and A.~Tonero, \emph{{Top-quark cross sections and
  distributions at approximate N3LO}},
  \href{https://doi.org/10.1103/PhysRevD.108.054012}{\emph{Phys. Rev. D}
  {\bfseries 108} (2023) 054012}
  [\href{https://arxiv.org/abs/2306.06166}{{\ttfamily 2306.06166}}].

\bibitem{Howarth:2020uaa}
J.~Howarth, \emph{{Elastic Potential: A proposal to discover elastic production
  of top quarks at the Large Hadron Collider}},
  \href{https://arxiv.org/abs/2008.04249}{{\ttfamily 2008.04249}}.

\bibitem{ATLAS:2018acq}
{The ATLAS collaboration}, \emph{{Measurements of differential cross sections
  of top quark pair production in association with jets in ${pp}$ collisions at
  $\sqrt{s}=13$ TeV using the ATLAS detector}},
  \href{https://doi.org/10.1007/JHEP10(2018)159}{\emph{JHEP} {\bfseries 10}
  (2018) 159} [\href{https://arxiv.org/abs/1802.06572}{{\ttfamily
  1802.06572}}].

\bibitem{Goncalves:2020saa}
V.~P. Gon\c{c}alves, D.~E. Martins, M.~S. Rangel and M.~Tasevsky, \emph{{Top
  quark pair production in the exclusive processes at the LHC}},
  \href{https://doi.org/10.1103/PhysRevD.102.074014}{\emph{Phys. Rev. D}
  {\bfseries 102} (2020) 074014}
  [\href{https://arxiv.org/abs/2007.04565}{{\ttfamily 2007.04565}}].

\bibitem{Martins:2022dfg}
D.~E. Martins, M.~Tasevsky and V.~P. Goncalves, \emph{{Challenging exclusive
  top quark pair production at low and high luminosity LHC}},
  \href{https://doi.org/10.1103/PhysRevD.105.114002}{\emph{Phys. Rev. D}
  {\bfseries 105} (2022) 114002}
  [\href{https://arxiv.org/abs/2202.01257}{{\ttfamily 2202.01257}}].

\bibitem{deFavereaudeJeneret:2009db}
J.~de~Favereau~de Jeneret, V.~Lemaitre, Y.~Liu, S.~Ovyn, T.~Pierzchala,
  K.~Piotrzkowski et~al., \emph{High energy photon interactions at the {LHC}},
  \href{https://arxiv.org/abs/0908.2020}{{\ttfamily 0908.2020}}.

\bibitem{Fayazbakhsh:2015xba}
S.~Fayazbakhsh, S.~T. Monfared and M.~Mohammadi~Najafabadi, \emph{Top quark
  anomalous electromagnetic couplings in photon-photon scattering at the
  {LHC}}, \href{https://doi.org/10.1103/PhysRevD.92.014006}{\emph{Phys. Rev. D}
  {\bfseries 92} (2015) 014006}
  [\href{https://arxiv.org/abs/1504.06695}{{\ttfamily 1504.06695}}].

\bibitem{dEnterria:2009cwl}
D.~d'Enterria and J.-P. Lansberg, \emph{{Study of Higgs boson production and
  its b anti-b decay in gamma-gamma processes in proton-nucleus collisions at
  the LHC}}, \href{https://doi.org/10.1103/PhysRevD.81.014004}{\emph{Phys. Rev.
  D} {\bfseries 81} (2010) 014004}
  [\href{https://arxiv.org/abs/0909.3047}{{\ttfamily 0909.3047}}].

\bibitem{Luszczak:2018dfi}
M.~\L{}uszczak, L.~Forthomme, W.~Sch\"afer and A.~Szczurek, \emph{{Production
  of $ t\overline{t} $ pairs via $\gamma\gamma$ fusion with photon transverse
  momenta and proton dissociation}},
  \href{https://doi.org/10.1007/JHEP02(2019)100}{\emph{JHEP} {\bfseries 02}
  (2019) 100} [\href{https://arxiv.org/abs/1810.12432}{{\ttfamily
  1810.12432}}].

\bibitem{CMS:2023naq}
{The CMS and TOTEM Collaborations}, \emph{{Search for central exclusive
  production of top quark pairs in proton-proton collisions at $\sqrt{s}$ = 13
  TeV with tagged protons}},
  \href{https://arxiv.org/abs/2310.11231}{{\ttfamily 2310.11231}}.

\bibitem{Sirunyan:2017ezt}
{The CMS Collaboration}, \emph{{Identification of heavy-flavour jets with the
  CMS detector in $pp$ collisions at 13~TeV}},
  \href{https://doi.org/10.1088/1748-0221/13/05/P05011}{\emph{JINST} {\bfseries
  13} (2018) P05011} [\href{https://arxiv.org/abs/1712.07158}{{\ttfamily
  1712.07158}}].

\bibitem{CMS:2021ncv}
{The CMS collaboration}, \emph{{The CMS Precision Proton Spectrometer at the
  HL-LHC -- Expression of Interest}},
  \href{https://arxiv.org/abs/2103.02752}{{\ttfamily 2103.02752}}.

\bibitem{Pitt:2023vpo}
M.~Pitt, \emph{{Physics at the HL-LHC with Proton Tagging}},
  \href{https://doi.org/10.5506/APhysPolBSupp.16.7-A12}{\emph{Acta Phys. Polon.
  Supp.} {\bfseries 16} (2023) 7}.

\bibitem{97898112801840018}
M.~Deile and M.~Ta\v{s}evsk\'y, \emph{High Luminosity Forward Physics}. WORLD
  SCIENTIFIC, 2023,
  \href{https://doi.org/10.1142/9789811280184\_0018}{10.1142/9789811280184\_0018}.

\bibitem{ATLAS:2015iqc}
{\scshape ATLAS} collaboration, \emph{{Search for single top-quark production
  via flavour-changing neutral currents at 8 TeV with the ATLAS detector}},
  \href{https://doi.org/10.1140/epjc/s10052-016-3876-4}{\emph{Eur. Phys. J. C}
  {\bfseries 76} (2016) 55} [\href{https://arxiv.org/abs/1509.00294}{{\ttfamily
  1509.00294}}].

\bibitem{Goldouzian:2016mrt}
R.~Goldouzian and B.~Clerbaux, \emph{{Photon initiated single top quark
  production via flavor-changing neutral currents at the LHC}},
  \href{https://doi.org/10.1103/PhysRevD.95.054014}{\emph{Phys. Rev. D}
  {\bfseries 95} (2017) 054014}
  [\href{https://arxiv.org/abs/1609.04838}{{\ttfamily 1609.04838}}].
  
\end{thebibliography}

\end{document}